\documentclass[referee]{raa}            

\usepackage{graphicx,times}             
\usepackage{natbib}
\usepackage{amssymb,amsmath}
\bibpunct{(}{)}{;}{a}{}{,}

\usepackage{txfonts}
\usepackage{xcolor}
\definecolor{forestgreen}{rgb}{0.10, 0.50, 0.10}
\renewcommand{\textbf}[1]{#1}

\usepackage[a4paper=true,pagebackref=true]{hyperref} 
\hypersetup{colorlinks = true, linkcolor = green, anchorcolor = red, citecolor = blue, filecolor = red, pagecolor = red, urlcolor = red}

\begin{document}
 
   \title{Generate Radioheliograph Image from SDO/AIA Data \\
   	with Machine Learning Method
\,$^*$
\footnotetext{$*$ Corresponding author: ChuanBing Wang,  \it{cbwang@ustc.edu.cn}}
}

   \volnopage{Vol.0 (20xx) No.0, 000--000}      
   \setcounter{page}{1}          

   \author{PeiJin Zhang
      \inst{1,2,3}
   \and ChuanBing Wang
      \inst{1,2,3}
   \and GuanShan Pu
      \inst{1,2,3}
   }

   \institute{CAS Key Laboratory of Geospace Environment,
   	School of Earth and Space Sciences, \\
   	University of Science and Technology of China,
   	Hefei, Anhui 230026, China; {\it cbwang@ustc.edu.cn}\\
        \and
             CAS  Center for the Excellence in Comparative Planetology, USTC, Hefei, Anhui 230026, China\\
        \and
             Anhui Mengcheng Geophysics National  Observation and Research Station, USTC, Mengcheng, Anhui 233500, China\\
\vs\no
   {\small Received~~20xx month day; accepted~~20xx~~month day}}

\abstract{ The radioheliograph image is essential for the study of solar short term activities and long term variations, while the continuity and granularity of radioheliograph data is not so ideal, due to the short visible time of the sun and the complex electron-magnetic environment near the ground-based radio telescope. In this work, we develop a multi-channel input single-channel output neural network, which can generate radioheliograph image in microwave band from the Extreme Ultra-violet (EUV) observation of the Atmospheric Imaging Assembly (AIA) on board the Solar Dynamic Observatory (SDO). The neural network is trained with nearly 8 years of data of Nobeyama Radioheliograph (NoRH) at 17\,GHz and SDO/AIA {from January 2011 to September 2018}. {The generated radioheliograph image is in good consistency with the well-calibrated NoRH observation. SDO/AIA provides solar atmosphere images in multiple EUV wavelengths every 12 seconds from space, so the present model can fill the vacancy of limited observation time of microwave radioheliograph, and support further study of the relationship between the microwave and EUV emission.}
}

   \authorrunning{Zhang et. al }            
   \titlerunning{Generate Radioheliograph Image with Machine Learning Method}  

   \maketitle

%
%

\section{Introduction}\label{sec1}
Microwave radioheliograph data is essential for the diagnose of solar atmosphere and the surveillance of solar activity \citep{shibasaki2013long,huang2009statistical,mei2017detection,tan2016diagnosing}. The quality and the continuity of the radioheliograph data is not as good as optical observation. The optical devices, for example, the Extrem Ultra-violet (EUV) observatory, can be launched into space, because of relatively small aperture comparing to the radioheliograph array \citep{lemen2011atmospheric,sandel2000extreme,wulser2004euvi}. The spacecraft can provide more continuous observation for the sun, compared with the ground-based observatory which cannot observe the sun for all-time due to the limited visible duration of the sun. Moreover, the ground-based radio devices are more liable to be affected by the complex electromagnetic environment, which will cause calibration problems. In this work, we intend to develop a machine-learning program which can learn the patterns of solar EUV images and the well-calibrated radioheliograph images, then produce the radioheliograph image from EUV data to provide full-time microwave heliograph.

{Conventionally, the modeling and reconstruction of radioheliograph image from EUV observation are based on the differential emission measure (DEM) method. Both the EUV radiation and radio emission come from the thermal electrons in the corona during non-flaring time. The corona electron density ($n_e$) and temperature ($T_e$) distribution can be obtained by DEM inversion of the multi-channel EUV emission \citep{pallavicini1981closed, ChengDEM2012}, then the radio brightness temperature can be derived from $n_e$ and $T_e$ distribution with a given emission mechanism. 
\cite{zhang2001reconciling} used thermal bremsstrahlung emission mechanism to predict the microwave heliograph image with the EUV data of EIT (Extreme-Ultraviolet Imaging Telescope) on-board SOHO (Solar and Heliospheric Observatory), and find that the predicted radio flux is systematically larger than that observed by a factor of 2.0. On the other hand,  \cite{alissandrakis2019modeling} modeled the sunspot-associated microwave emission based on the gyro-resonance emission mechanism with potential extrapolations of the photospheric magnetic field, in which the DEM was inverted from the EUV image of Atmospheric Imaging Assembly (AIA) on-board the Solar Dynamics Observatory (SDO) \citep{lemen2011atmospheric}. More recently, \cite{li2020synthesising} find that the predicted radio flux is closer to the observations in the case that includes the contribution of plasma with temperatures above 3\,MK than in the case of only involving low temperature plasma, and confirmed the thermal origin of the quiet corona radio emission. The predicted value of the DEM method depends on the physics model, including the derivation of $n_e$, $T_e$ and magnetic field, and the emission mechanism.} 

{The observed radio flux in the line of sight is a convoluted result of the wave excitation and the wave propagation process \citep{shibasaki2011radio, tan2015study}. The brightness temperature at a given position is affected by both local conditions and the large-scale coronal structures. This makes some difficult to the modeling of radio emission from EUV data with a pixel to pixel translation (such as in the DEM method), while the machine-learning method with a convolutional neural network (CNN) has the advantage of connecting areas to pixels.}

The machine-learning method has recently been applied to solar physics \citep{BobraMason2019}. { \cite{ma2017multimodal} used a neural network based on the multimodal learning architecture to classify the existence of the radio burst. \textbf{\cite{li2013solar} used a multi-layer model to predict the solar flare based on sequential sunspot data.} \cite{xu2019lstm} used Long short term memory (LSTM) network to classify multiple types of the solar radio spectrum.}  As a large set of image data, SDO data is suitable for various purposes of machine learning work. Neural networks like CNN and the generative adversarial networks (GAN) can be used for data generation and competition. For examples, \cite{szenicer2019deep} used a combined CNN to produce the EUV irradiance map from SDO/AIA image, \cite{kim2019naturesolar} applied GAN to generate the magnet flux distribution of the Sun from SDO/AIA image, and \textbf{\cite{xu2020solar} used the GAN for the de-convolution of the solar radio image}.

In this work, we {proposed a model for radioheliograph reconstruction from SDO/AIA multi-channel EUV images, using the machine-learning method based on a multi-channel input single-channel  output  neural  network}. In Section 2, {the dataset,} the architecture and training process of the neural network is elaborated. In Section 3, we present \textbf{the statistical results and} some representative cases to show the reliability of the trained neural network. In section 4, we summarize the result and discuss the meaning of the result and further usage of this method.

\section{Method}\label{sec2}
\subsection{Dataset}
In this work, the microwave image data we use is from the Nobeyama Radioheliograph   (\href{https://solar.nro.nao.ac.jp/norh/html/daily/}{NoRH}) \citep{nakajima1994nobeyama}. The EUV image data we use is provided by SDO/AIA \citep{lemen2011atmospheric}, which is downloaded with sunpy \citep{mumford2015sunpy} from Virtual Solar Observatory (\href{https://sdac.virtualsolar.org/cgi/search}{VSO}). The dataset we used is the 17\,GHz radioheliograph image at noon of Nobeyama local time from January 2011 to September 2018, and the SDO/AIA image of five wavelengths: 171, 193, 211, 304, 335 $\AA$ for the corresponding time. The original size of the EUV image from SDO/AIA is (4096, 4096)\,pixels, while the original size of radio heliograph image from  NoRH is (512, 512)\, pixels. For the convenience of further training, the original data is feed to the following prepossessing steps:
\begin{enumerate}
	\item Shift the solar center to the center of the image.
	\item Crop the image to 1.1 times of the solar radius.
	\item Re-sample the EUV image into the size of (512, 512) pixels.
	\item Modify the NaN (not a number) points and minus values in the data to zero.
	\item Normalize the brightness temperature with $10^4$\,K to avoid the byte overflow of float-type number during the training.
	\item Save as Numpy form (.npy) file format for faster load.
	\item Manually select and mark the corrupted or {not well-calibrated data frames}, which would be sheltered from the training.
\end{enumerate}
The final size of the total dataset is about 18 GB with 2622 data frames, and the data is stored in RAMDisk to accelerate the data loading in the training and testing process. 

\subsection{Neural Network Architecture}

\begin{figure}[ht]%
	\centering
	\includegraphics[width=1.\linewidth]{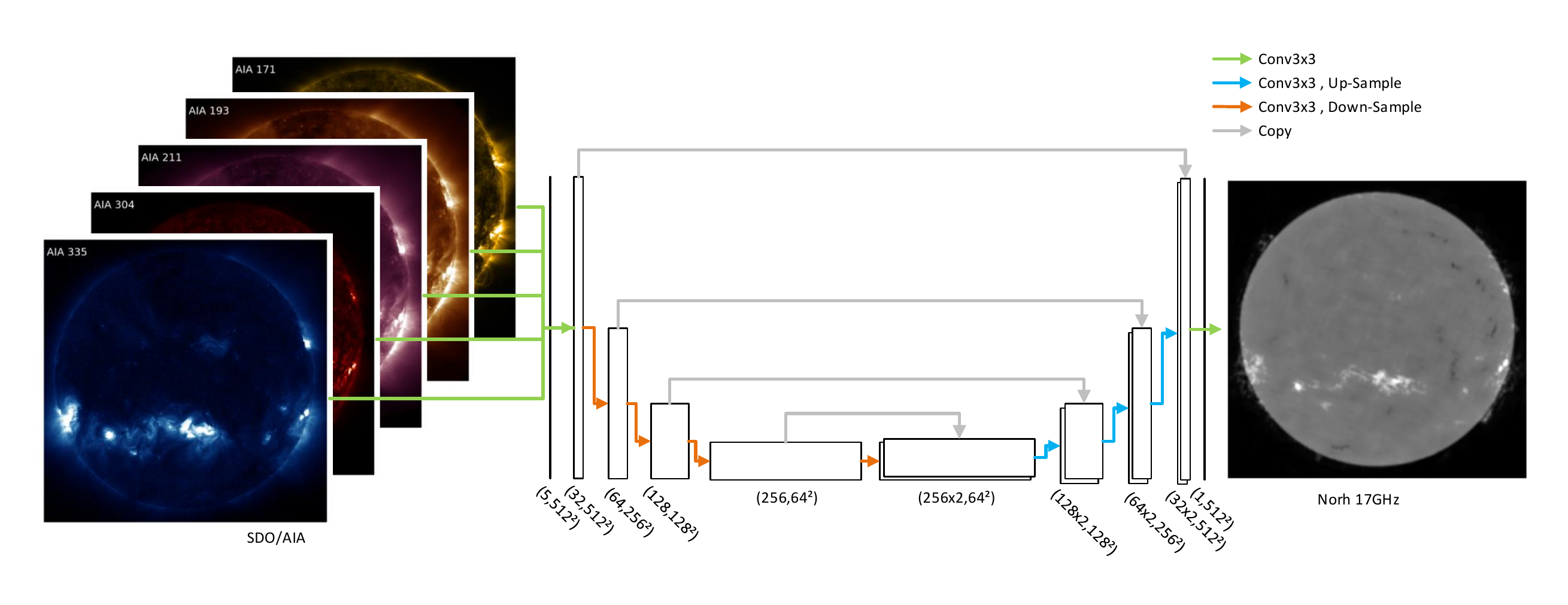}%
	\caption{The architecture of the neural network, where the arrows represents the data flow direction, the white blocks represents the nodes in the data flow, and the size of the each node is marked beneath the block.} %
	\label{fig:1}%
\end{figure}

We used a multi-input single-output four level U-Net (abbreviated as MISO-UNet) to produce the microwave heliograph data. The detailed architecture is shown in Figure (\ref{fig:1}). This MISO-UNet has $3.35\times 10^6$ parameters, the size of the trained parameter-set file is 12.79\,MB. The loss-function is the Mean Square Error (MSE) between the observed NoRH image (labeled as OBS)  and the neural network generated image (labeled as GEN):
\begin{equation}
Loss = {\frac{\sum_{i} (I_{OBS,i}-I_{GEN,i})^2}{N_{pix}}},
\end{equation}
where $I_{OBS,i}$ and $I_{GEN,i}$ are the normalized brightness temperature of the $i$th pixel in the observed image and the generated image, respectively. The value of brightness temperature is normalized with $10^4$\,K. $N_{pix}$ is the total number of pixel in the training data. The neural network is implemented with PyTorch \citep{paszke2017automatic,paszke2019pytorch}, an open source machine learning library. The source code of the data processing and 
neural network is available online \footnote{\url{https://github.com/Pjer-zhang/NorhBot}}.

\subsection{Training}
We \textbf{randomly} selected out \textbf{235 frames} as the test set, which are sheltered from the training process. The rest \textbf{2387} frames of EUV-radio data is used for the training. The training of 8000 epochs took \textbf{94} hours on a single node with four Nvidia Titan Xp GPUs.  

The image pattern of radio heliograph is highly related to observation time or the sequence index. There are more active regions at solar high year, while the radio heliograph of solar low year is more similar to pure disk. As a result, the sequence index of the data frame also directly contains the information of radio heliograph image. For a better learning of the relation between EUV and radio heliograph, we need to intermingle the dataset, so that the neural network is not fed with the information of data order.  During the training process, the data frames are shuffled in each epoch. 

The converged value of the loss-function is about \textbf{ $7\times10^{-4}$}, corresponding to about \textbf{300K} in brightness temperature.

\section{Results}
After the training process, the trained parameter-set can be loaded into the MISO-UNet model for the production of radio heliograph image. We use the data frames from the test set to verify the reliability of the trained model.  

\begin{figure}[ht]
	\centering
	{\includegraphics[width = 0.33\linewidth]{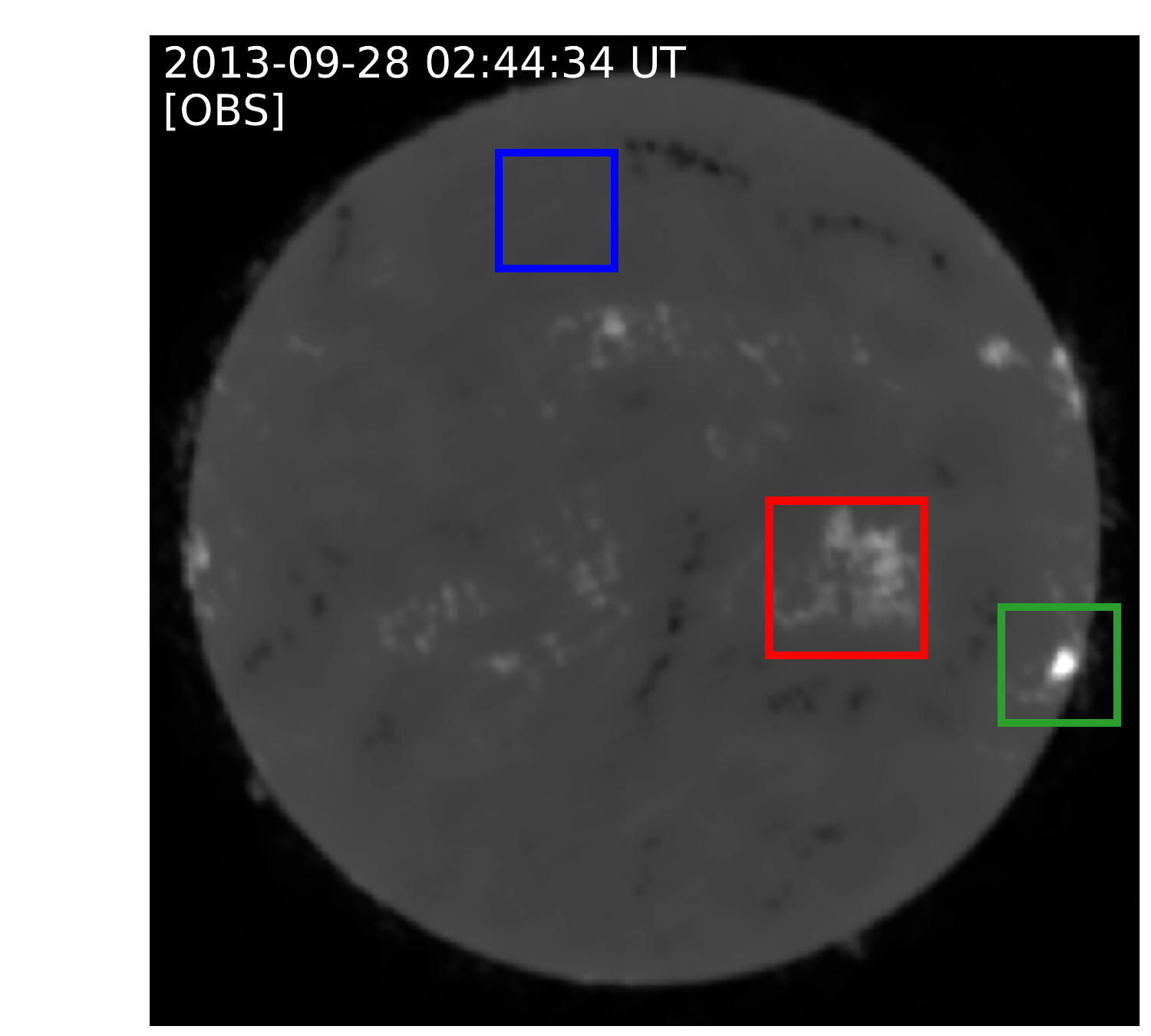}}
	{\includegraphics[width = 0.33\linewidth]{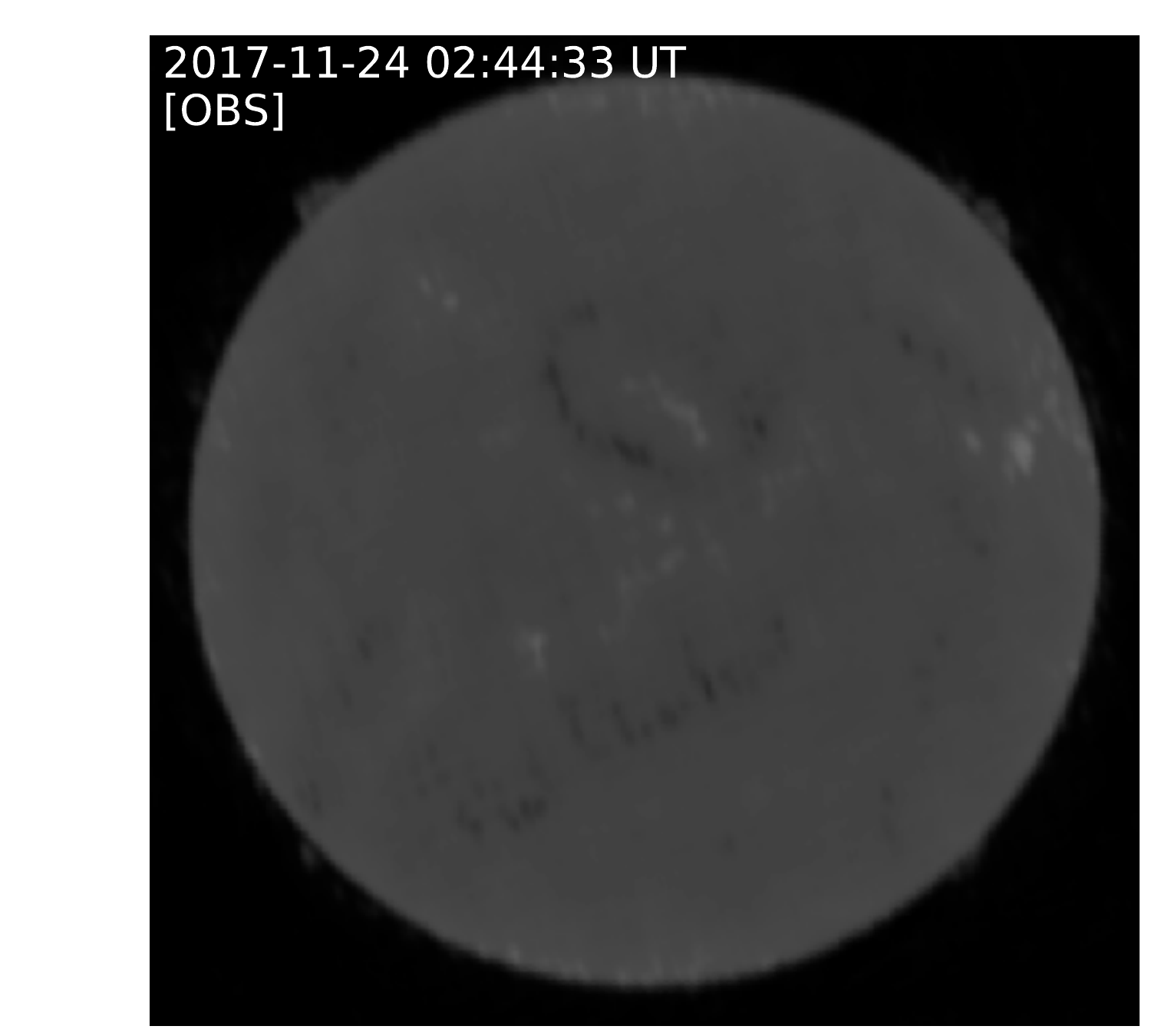}}
	{\includegraphics[width = 0.33\linewidth]{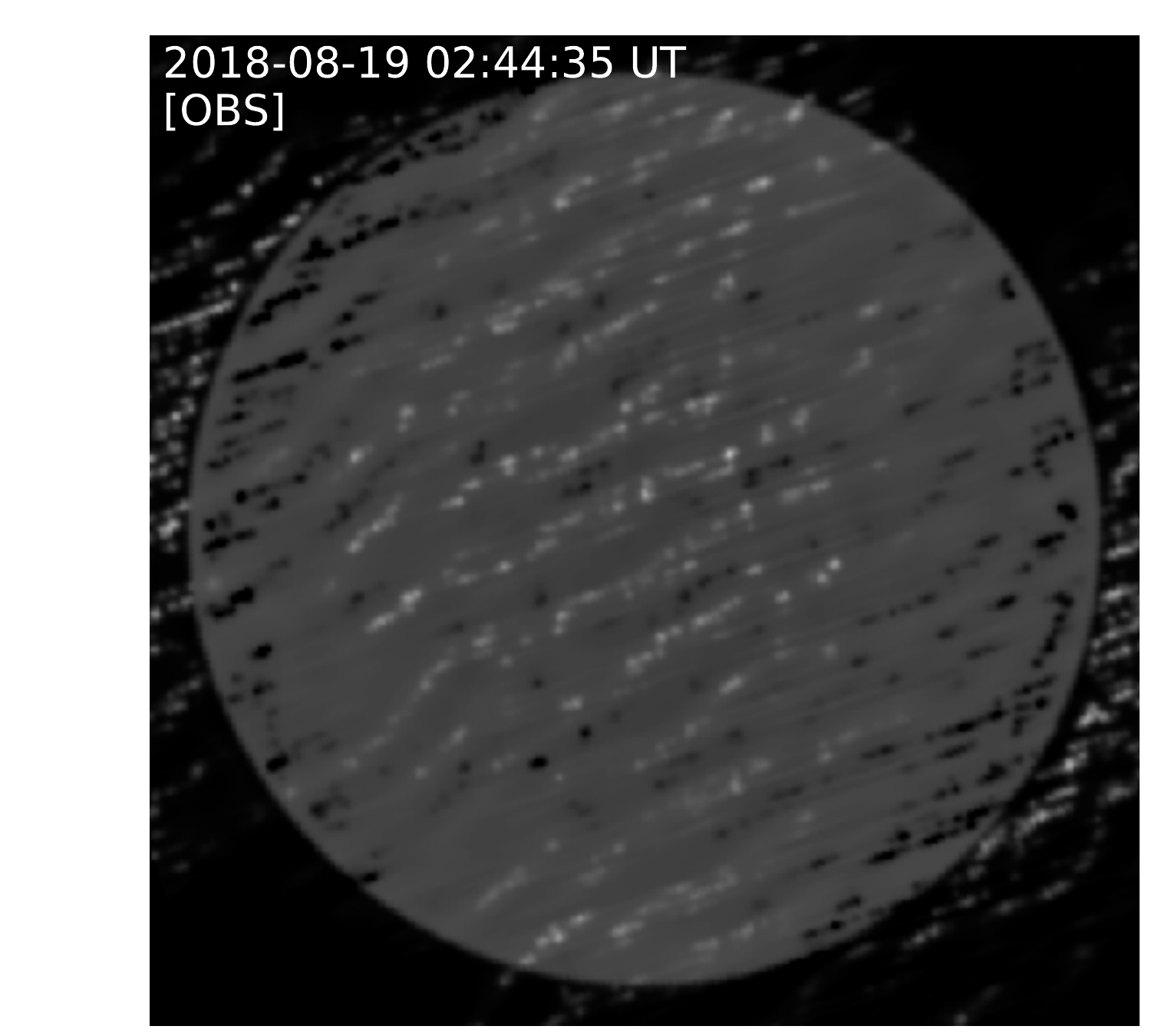}}
	{\includegraphics[width = 0.33\linewidth]{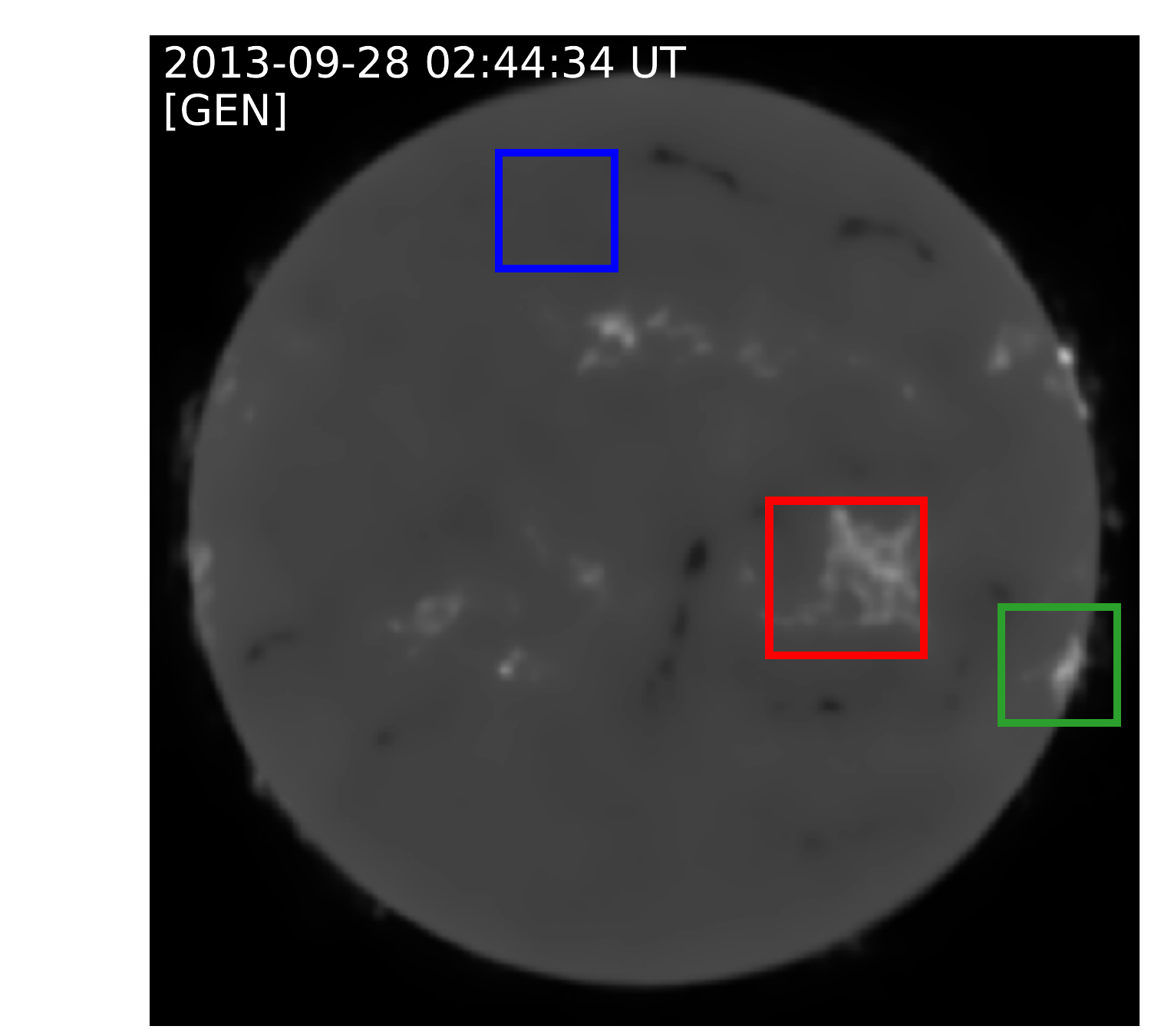}}
	{\includegraphics[width = 0.33\linewidth]{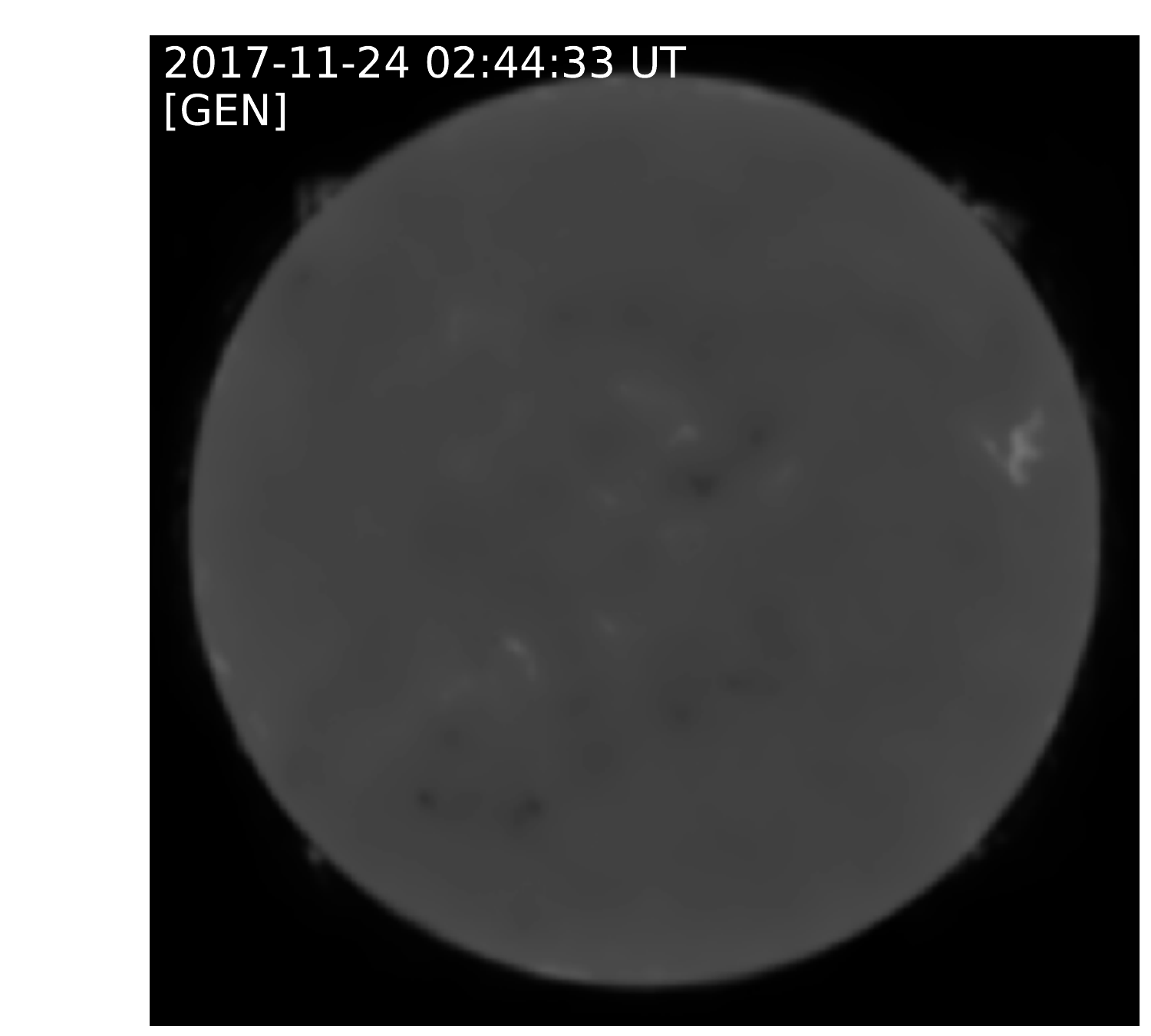}}
	{\includegraphics[width = 0.33\linewidth]{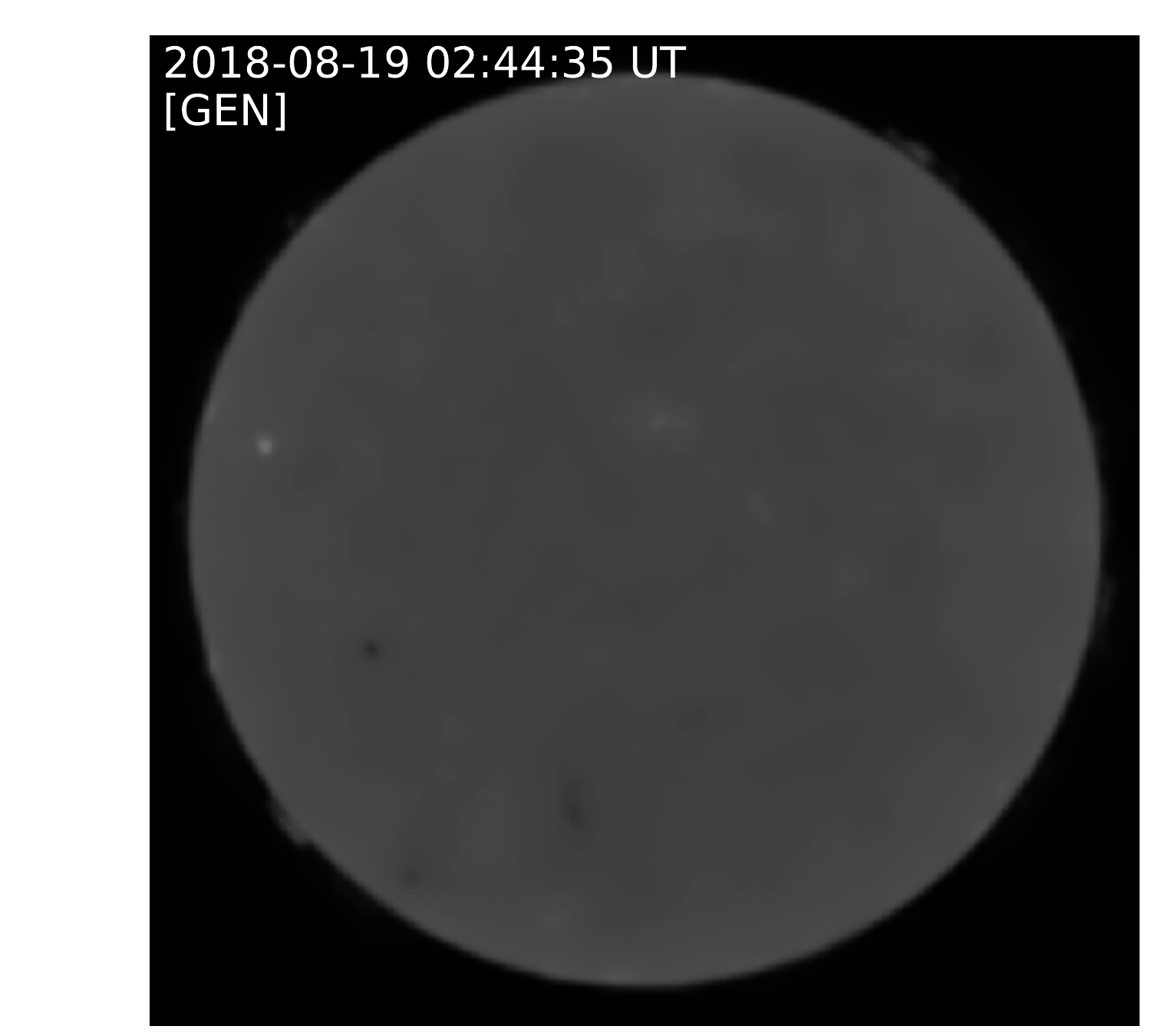}}
	{\includegraphics[width = 0.33\linewidth]{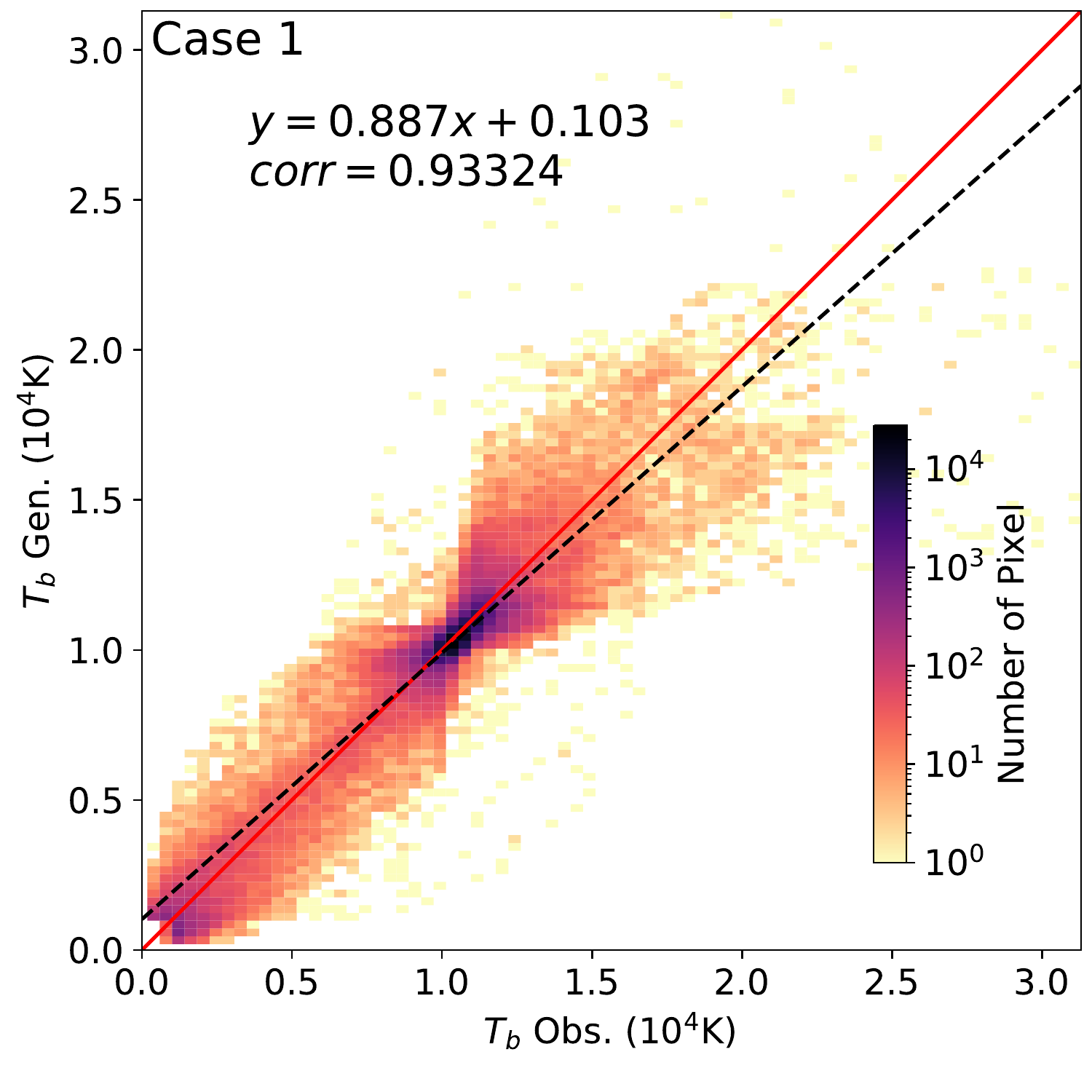}}
	{\includegraphics[width = 0.33\linewidth]{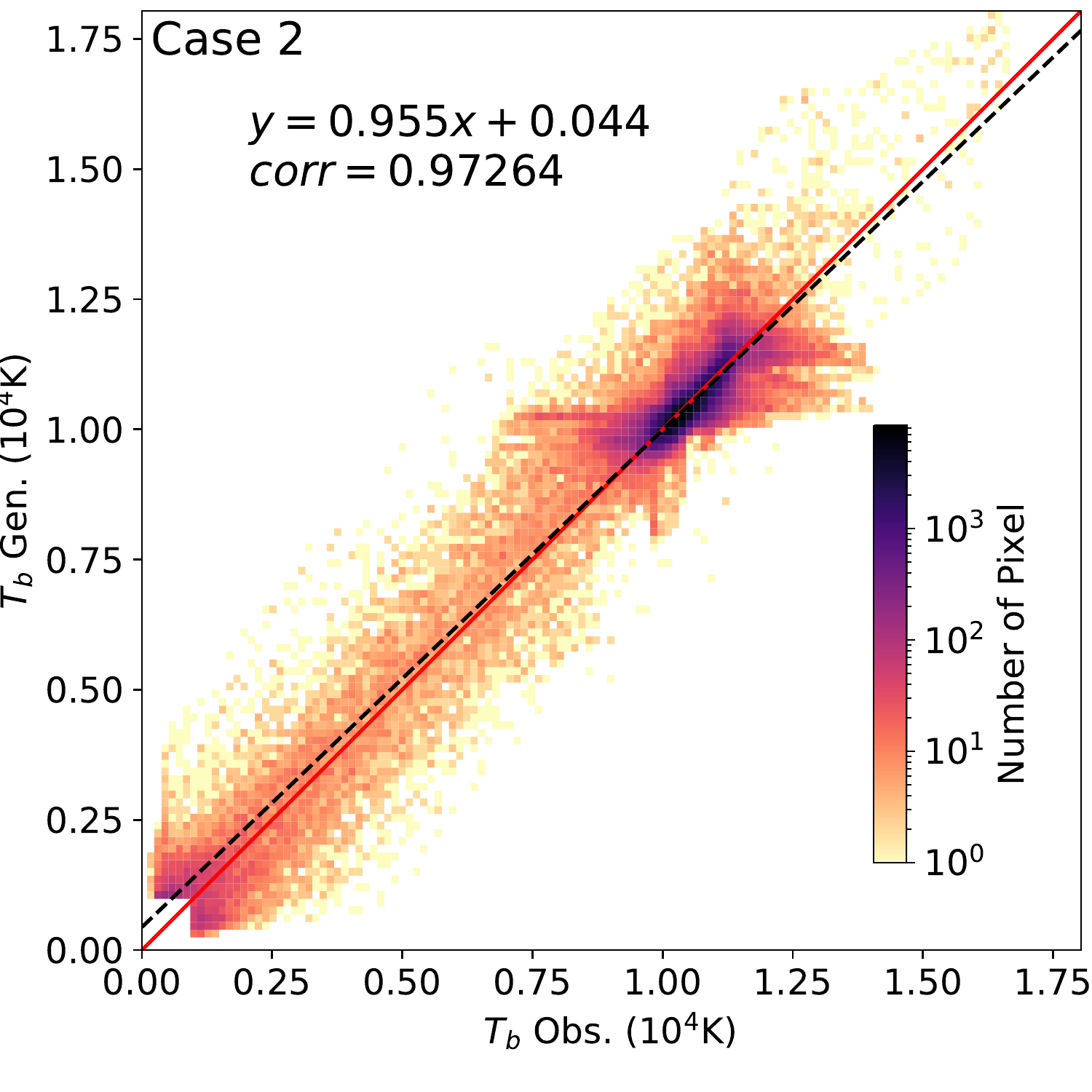}}
	{\includegraphics[width = 0.33\linewidth]{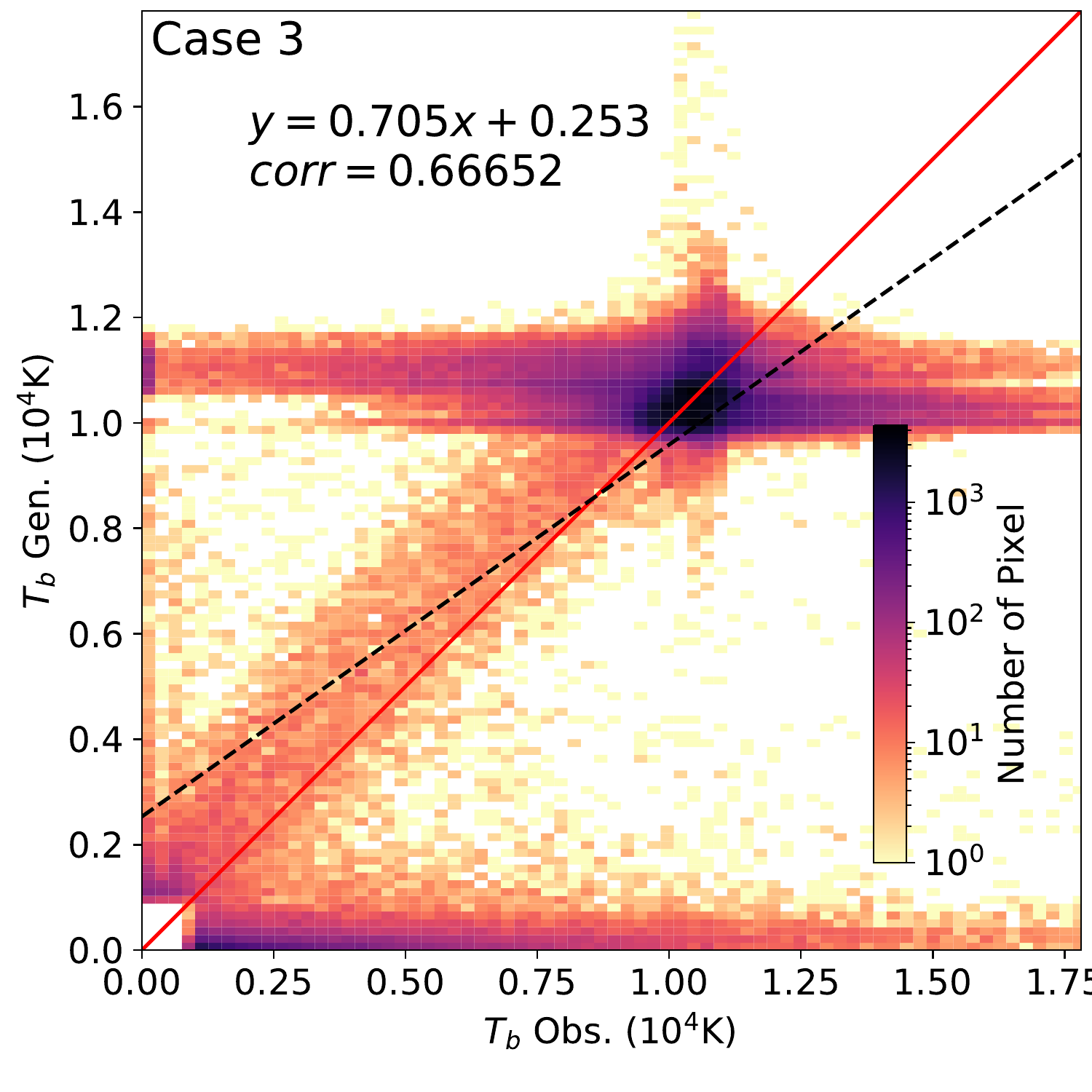}}
	\caption{Three cases of comparing the observed image and neural network generated image and the corresponding statics. These three cases are all not included in the training dataset. The date-time of the image is marked on the right-top corner of the image. The top row is the observed image, the middle row is the generated image using 5-channel EUV data. The bottom row shows the correlation of the observed data and the generated data, where the $x$-axes is the observed brightness temperature, the $y$-axes is the generated brightness temperature, the color indicates the number of pixels with corresponding generated and observed brightness temperature, the red line marks the reference line of $x$ equals $y$, \textbf{the black dashed line indicates the linear fit result $y=ax+b$}. The linear fit result and the correlation between the observed and generated flux are labeled in each panel of the bottom row. }
	\label{fig:main}
\end{figure}

\begin{figure}[ht]
	\centering
	{\includegraphics[width = 0.75\linewidth]{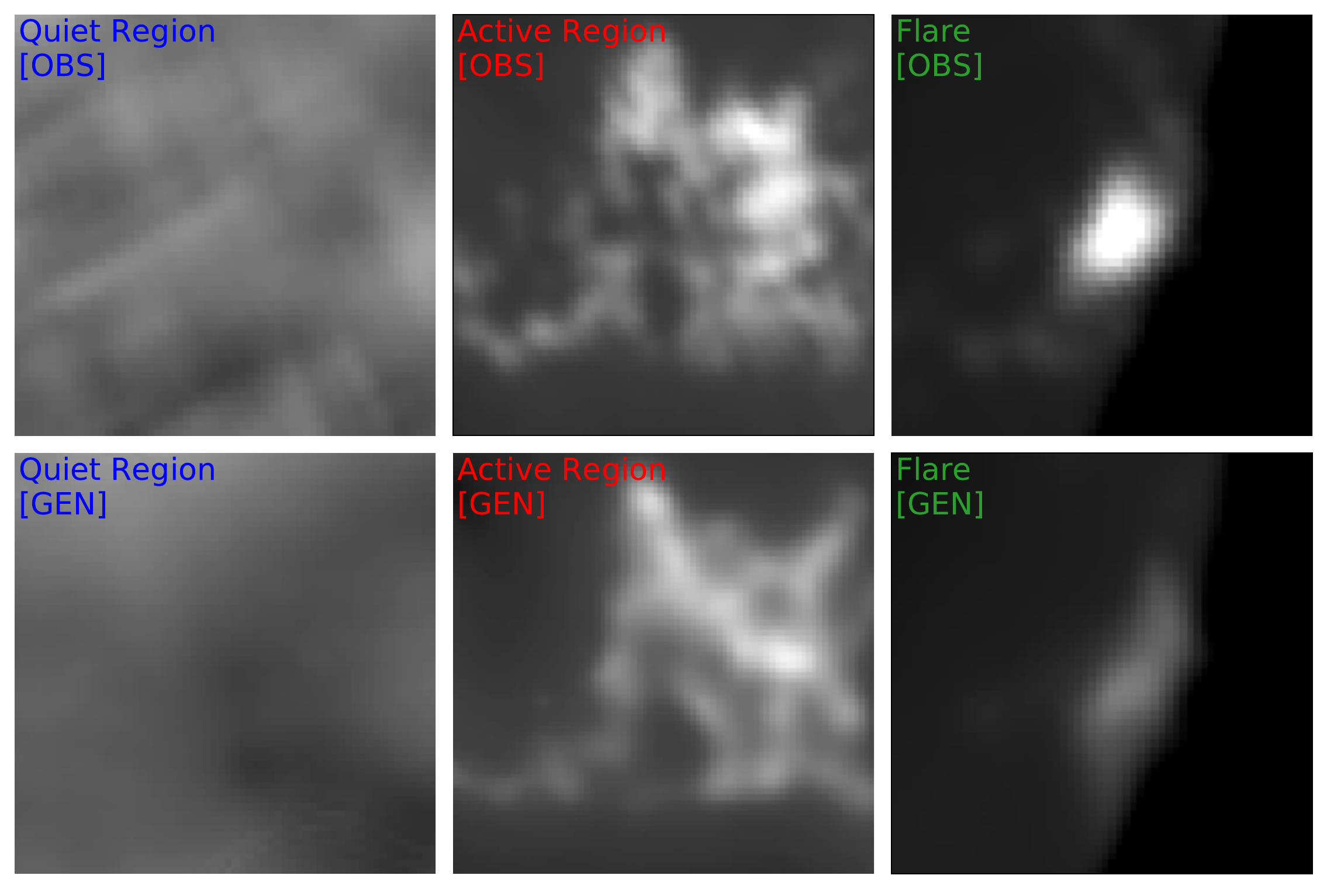}}
	\caption{The zoom-in comparison of the observed and MISO-UNet generated image. \textbf{These three columns from left to right present the results for a quiet region, an active region, and a flare, respectively. These there regions are marked as blue, red and green box in Figure \ref{fig:main}}, respectively. Note that the color-scale is re-adjusted to show the difference.}
	\label{fig:zoom}
\end{figure}

\begin{figure}[ht]
	\centering
	{\includegraphics[width = 0.6\linewidth]{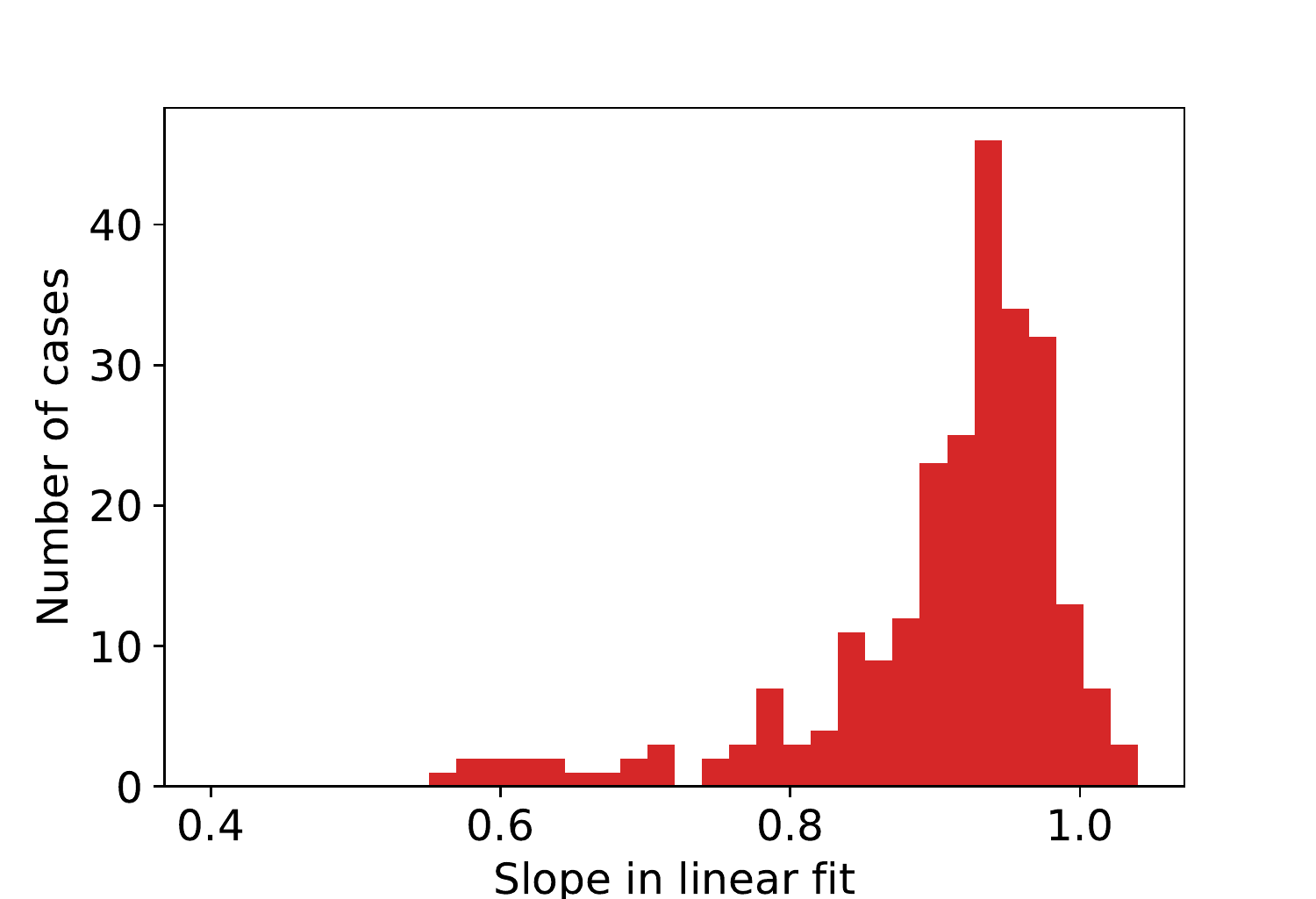}
	
	\includegraphics[width = 0.6\linewidth]{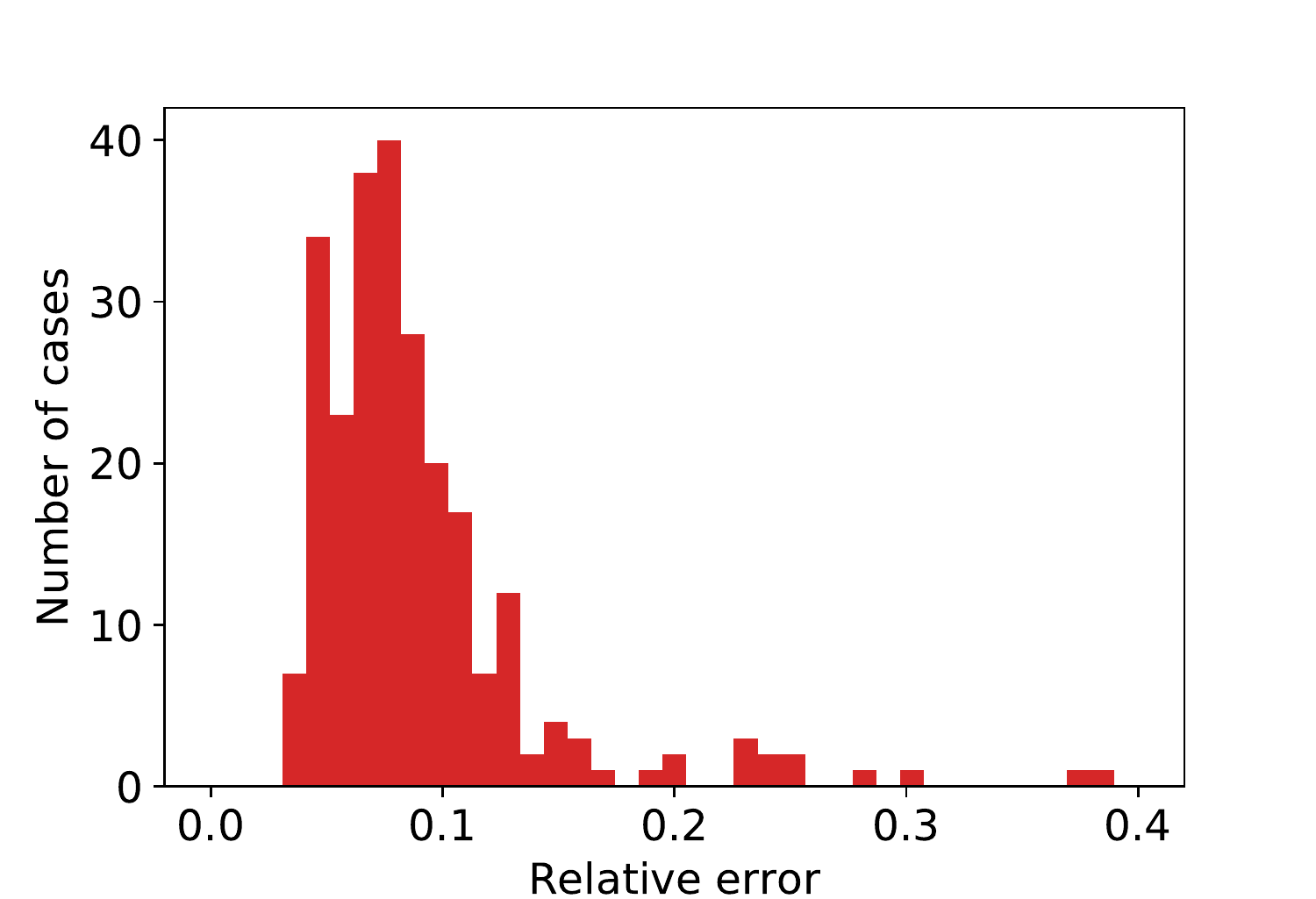}
	
	\includegraphics[width = 0.6\linewidth]{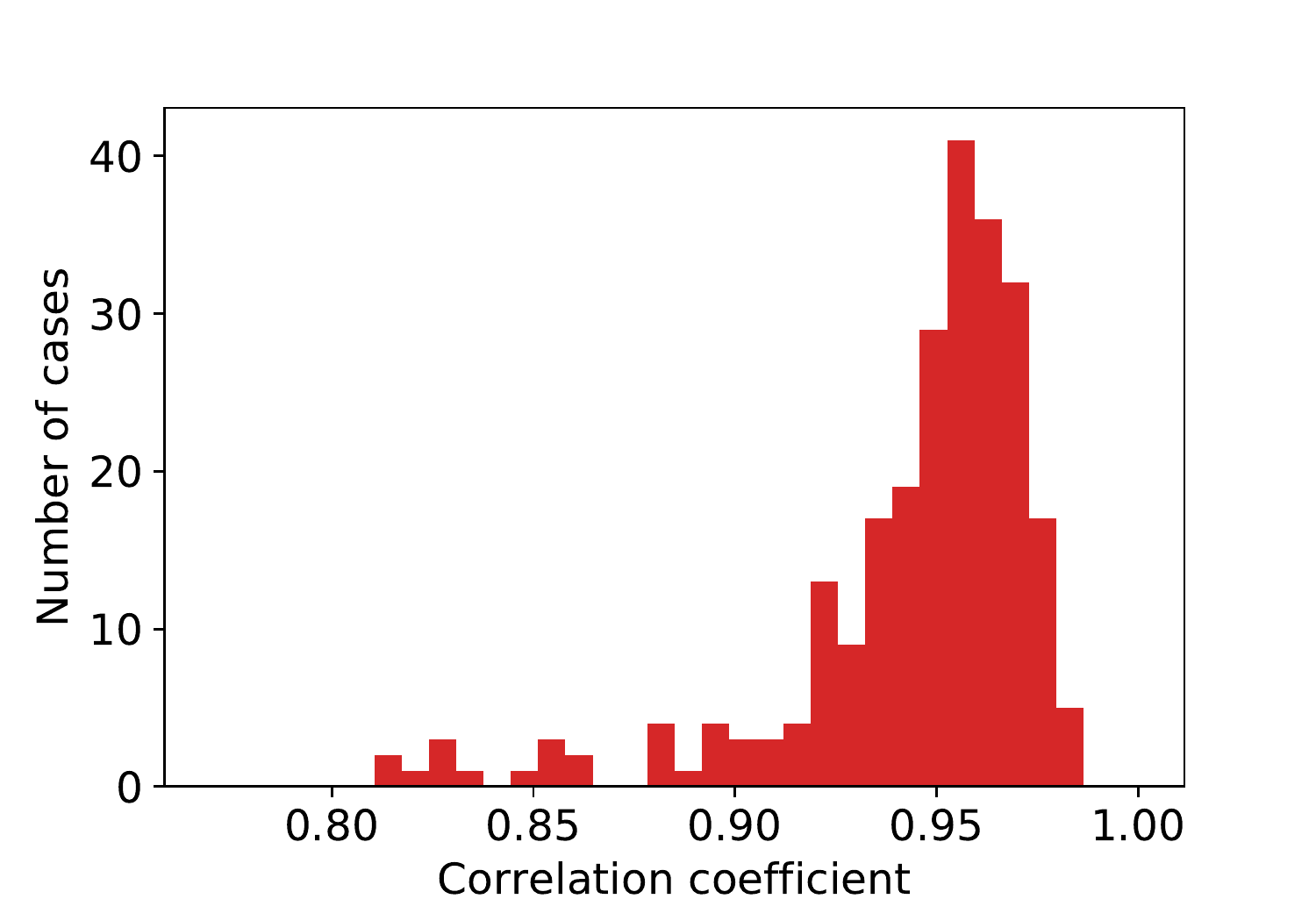}
	
	}
	\caption{{The histogram of the linear fit slope, the relative error, and the correlation coefficient between GEN and OBS for the test cases. The relative error is defined as $\rm{RMSE}\it{(I_{GEN},I_{OBS})}/\rm{average} \it{(I_{OBS})}$}}.
	\label{fig:hist}
\end{figure}

It is found that the neural network generated image is well consistent with the observed one. Figure \ref{fig:main} shows three representative cases from the test-set as examples for the comparison of the observed radio images and the neural network generated ones in detail. From left to right, the first column is a case in solar high year, where multiple active regions can be seen on the solar surface.  The second column represents a case in solar low year. The third column  represents a case of bad observation, mainly from the  radio-frequency interference of the complex electromagnetic environments.

The linear fit result of Case 1 and Case 2 is close to $y=x$, and the correlation value is close to 1.0. \textbf{To exclude the dark area outside the solar disk in NoRH observation, we only consider the pixels with brightness temperature greater than 1000 K in both GEN and OBS in the statistical investigation.} The results indicate that the flux intensity of the generated image and the observed image are well correlated except the observation is corrupt (as case 3 shown in the third column of Figure \ref{fig:main}). Moreover, from the first case, \textbf{we selected three regions to show the detailed comparison of an active region, a quiet region, and a flare, which are marked as blue, red, and green box in the first column of Figure \ref{fig:main}}. Figure \ref{fig:zoom} shows the detailed zoom in of the comparison. From the comparison of the GEN and OBS of the quiet region, one can see regular spaced stripe structures in OBS, which are not visible in GEN. For the structure and shape of the active region, GEN and OBS are generally consistent with each other. \textbf{However, the flare region is not well reconstructed by this method, and the generated brightness temperature is much lower than the observed value as shown in the third column of Figure \ref{fig:zoom}.}

\textbf{Figure \ref{fig:hist} shows the histogram distribution of the linear fit slope $a$, the relative error ($re$), and the correlation coefficient ($cc$) between GEN and OBS for the test data set.} \textbf{ It is found that ninety percent of the linear fit slope falls in the range  \textbf{$0.80<a<1.02$, with an average value of 0.91.} The average value of $re$  is 0.09, and for 90\% of the test frames, $re$ is less than 0.1. The average value of $cc$ is 0.94, and for 90\% of the test frames, $cc$ is above $0.91$. By inspecting every frame of the test cases, we find that the frames with $a<0.8$, $cc<0.9$, or $re>0.1$ are mostly embedded with a flare-like bright region.}


\section{Conclusion and Discussion}

This work proposed a model to reconstruct radioheliograph image from EUV image with a machine learning method based on MISO-UNet. The model generated radio image is \textbf{in good} consistent with the NoRH observed image at 17 \,GHz, which indicates that the EUV emission and the radio emission has a strong inner relation. \textbf{We inspected the image details of the generated images compared with the observation. As shown in Figure \ref{fig:zoom}, for quiet region there are regular spaced strips in the observed image, which is not appear in the generated images. We suppose the reason is that, the regular strips in the observation may due to the instrumental artifacts, which is arbitrary for different image. The training process learns the pattern between the EUV and radioheliograph image, the arbitrary information cannot be assimilated by the model. The structures of the active region is well reconstructed. The flux of the flare region is not well reconstructed. On one hand, this is due to the unbalanced sample distribution of the data-set, since the flare is rare compared to the active region and quiet sun. On the other hand, the radio emission at flare time is mainly produced by non-thermal electrons, while the non-flare emission are coming from the thermal electrons. The bad prediction result of the flare region may be improved by including more data-frames with flare in the future work.}
Comparing with the DEM method, we can have higher precision of brightness temperature prediction with the machine learning method. \textbf{The correlation of machine learning prediction is above 0.8 with an average value of 0.94 for the test cases, and the linear fit for the flux intensity of OBS and GEN is also much closer to $y=x$ than the DEM prediction results \citep{zhang2001reconciling,li2020synthesising}.} This may be partially due to the uncertainty of the DEM inversion and the emission mechanism assumed in the prediction. Either thermal bremsstrahlung emission or gyro-resonance radiation is assumed in the modeling with DEM method, while the quiet microwave emission is a mixture of the two mechanisms \citep{shibasaki2011radio}.

In addition, the DEM method of the radio flux estimation is a pixel-to-pixel operation, the final brightness temperature only considers the EUV flux of the pixel with the same coordinate in the sky plane. While in the MISO-UNet, the final brightness temperature value of any pixel considers not only the influence of the pixel with the same coordinate, but also the effects of surrounding large scale structures. The large and small scale features of all EUV channels spread though the neural  network to the final brightness temperature by convolution operations. {In fact, the observed radio flux is a convoluted results of the wave excitation and propagation process.} The brightness temperature of a given position is affected by both local conditions and the large-scale coronal structures, such as coronal holes, loop systems, and streamers etc \citep{shibasaki2011radio}. It is important to consider the nearby structures when deriving the observed radio emission of a given point. The MISO-UNet is suitable for the prediction of microwave emission map, since it has the advantage of connecting areas to pixels. {However, it needs to note that the machine learning method can only predict the data of the trained frequency which is fed to the training process, and the  DEM method can calculate the brightness temperature of the quiet Sun at different radio frequencies.}

{Finally, NoRH was shut down on March 31, 2020 after continuously observing the full sun for about eight hours every day in the past three decades. The trained MISO-UNet can produce radio heliograph image when SDO/AIA data is available. The present work is not only helpful to correct the bad observation data of NoRH due to electromagnetic interference etc, but also may potentially provide a virtual interferometric microwave observation from SDO/AIA data in the future. This method is also useful for the modeling and image reconstruction of radio observation in other wavelengths. The results can support further study of the relationship between the microwave and EUV emission.}


\section*{Acknowledgements}
{We thank the
	Nobeyama Radioheliograph (NoRH) and the SDO/AIA science team. NoRH is operated by the NAOJ/Nobeyama Solar Radio Observatory AJOS, and SDO/AIA is developed and operated by NASA.  This work was supported by the National Natural Science Foundation of China (Grant Nos. 41974199 and 41574167) and the B-type Strategic Priority Program of the Chinese Academy of Sciences (XDB41000000). We would like to acknowledge Drs. Masumi Shimojo and Satoshi Masuda from NAOJ, Dr. Tan Baolin from NAOC for help in delivering the NoRH data.}


\bibliographystyle{abbrv}
\bibliography{cite}

\end{document}